\pdfoutput=1
\documentclass[%
 amsmath,amssymb,
 aps,
]{revtex4}

\usepackage{graphicx}% Include figure files
\usepackage{dcolumn}% Align table columns on decimal point
\usepackage{bm}% bold math
\usepackage{amsmath}
\usepackage[english]{babel}
\usepackage{graphicx,color}
\usepackage{hyperref}

\newcommand{\Z}[2]{\ensuremath{Z^{#1\rightarrow #2}}}
\newcommand{\mess}[3]{\ensuremath{#1^{#2 \rightarrow #3}}}
\newcommand{\neigMeta}[1][e]{\ensuremath{\partial #1^M}}
\newcommand{\neigRea}[1][a]{\ensuremath{\partial #1^R}}

\def\neigGroupI{j \in \partial a^R_{i}}
\def\neigOppositeI{j \in \partial a^R_{\neg i}}
\def\neigIn{j \in \partial a^R_{in}}
\def\neigOut{j \in \partial a^R_{out}}

\newcommand{\avg}[1]{\left\langle{#1}\right\rangle}

\newcommand{\inn}{{\rm in}}
\newcommand{\outt}{{\rm out}}

\begin{document}

%\preprint{APS/123-QED}

\title{Boolean constraint satisfaction problems for reaction networks}% Force line breaks with \\

\author{
A. Seganti$^{1}$, 
A. De Martino$^{1,2,3}$ and
F. Ricci-Tersenghi$^{1,2,4}$
}
\affiliation{%
$^1$ Dipartimento di Fisica, Sapienza Universit\`a di Roma, p.le A.~Moro 2, 00185 Rome (Italy)\\
$^2$ IPCF-CNR, UOS di Roma, Dip.~Fisica, Sapienza Universit\`a di Roma (Italy)\\
$^3$ Center for Life Nano Science@Sapienza, Istituto Italiano di Tecnologia, Viale Regina Elena 291, 00161 Roma (Italy)\\
$^4$ INFN -- sezione di Roma1, Dip.~Fisica, Sapienza Universit\`a di Roma (Italy)
}%

%\date{~}% It is always \today, today,

\begin{abstract}
We define and study a class of (random) Boolean constraint satisfaction problems representing minimal feasibility constraints for networks of chemical reactions. The constraints we consider encode, respectively, for  hard mass-balance conditions (where the consumption and production fluxes of each chemical species are matched) and for  soft mass-balance conditions (where a net production of compounds is in principle allowed). We solve these constraint satisfaction problems under the Bethe approximation and derive the corresponding Belief Propagation equations, that involve 8 different messages. The statistical properties of ensembles of random problems are studied via the population dynamics methods. By varying a chemical potential attached to the activity of reactions, we find first order transitions and strong hysteresis, suggesting a non-trivial structure in the space of feasible solutions. 
\end{abstract}

\pacs{Valid PACS appear here}% PACS, the Physics and Astronomy
\maketitle 

\tableofcontents

\section{Introduction}

Biological networks map out the complex set of interactions that may occur among different units (genes, proteins, signalling molecules, enzymes, etc.) in cells \cite{Barabasi_2004,Tkacik_Bialek_2007}. Their structure is thought to reflect, at least in part, the specific physiologic function(s) they are meant to control and, with the topology-to-function mapping mostly still unclear, many important questions can be formulated about the optimality, robustness and evolvability of these systems \cite{Barabasi,Palsson_Theo,Alon,Crombach_2008,Berkhout_2012,Wagner_2008,Wagner_2005,Ciliberti_2007}. On the other hand,  functional control in cells is achieved through the physical dynamics that takes place {\it on} the networks, which is usually much more complicated than the network structure by itself would suggest. To make an example, consider transcriptional regulatory networks. While their structure only encodes for the possible protein-DNA interactions by which the transcription of RNA can be turned on or off, regulation results from the reciprocal adjustment of transcriptional activity and protein levels. This process however involves a variety of regulated steps, like DNA-binding and unbinding events by multiple proteins (possibly preceded by the formation of protein complexes), RNA polymerization (by specifically recruited molecular machinery) and transport, post-transcriptional modification events and, finally, translation. Each node in this network therefore lumps together a number of molecular species and elementary processes, all of which are subject to noise. In such a complex interacting environment, the overall patterns of activity may be hard to uncover even if one is only interested in steady states. 

It is tempting, then, to implement a coarse-grained approach and explore the possibility of characterizing the operation of biological networks through simpler, perhaps Boolean, dynamical rules or, at an even more basic level, through elementary feasibility constraints \cite{Kauffman,Leone_Zecchina,Martin,MartinII,Francois,Samal}. This type of scheme is especially suited (a) to identify robust attractors of the dynamics and/or groups of nodes that are likely to behave in a highly correlated way (viz. the emergence of network motifs discussed in \cite{Martin}), and (b) to evaluate `degrees of activity' for the different nodes, by which one may, for example, guide more refined techniques that simulate the full dynamics of the system towards physiologically relevant states. Besides, from a purely theoretical viewpoint, very often the problems thus defined present phase structures and algorithmic challenges that suffice by themselves to attract a considerable statistical mechanical effort (see e.g. \cite{corre,hamed}).

In this paper we define and study a Boolean constraint-satisfaction problem (CSP) designed to represent minimal operational and stability requirements for the non-equilibrium steady states (NESS) of biochemical reaction networks, like the metabolic networks that relate enzymes to the substrates and products of the reactions they catalyze in any given cell type \cite{Palsson_book,Heinrich_1996_book}. In essence, we shall enforce feasibility constraints that link enzyme activity to substrate and product availability, and vice-versa, similarly to the approach defined in \cite{Ebenhoh,Handorf_2007,Ebenhoh_2004}. From a physical viewpoint, the model describes, in different limits, different types of NESS, and therefore different physiological scenarios. The corresponding CSPs, on the other hand, turn out to be of a novel type, requiring {\it ad hoc} message-passing methods to be analyzed in detail. 

This article focuses on the properties of CSPs defined on ensembles of artificial (random) reaction networks -- introduced in Section II -- which will be studied by the statistical physics tools sketched, together with the corresponding results, in Section III (and fully exposed in the Appendix). A subsequent work will be concerned with the analysis of solutions for single random networks and real metabolic networks.

\section{Problem statement}

\subsection{Random Reaction Networks}

We define a random reaction network (RRN) to be a bipartite random graph with two types of nodes, representing respectively chemical species (or \textit{metabolites}) and enzymes (or \textit{reactions}). We shall denote by $N$ and $M$, respectively, the number of reactions and that of metabolites. Both $N$ and $M$ will be taken to be large, i.e. $N,M\gg 1$. For sakes of simplicity, we shall assume here that each reaction has a well defined operational direction, so that  the bipartite graph is directed. Its topology will be encoded in an adjacency matrix $\widehat{\xi}$, with entries $\xi_i^m=1$ if reaction $i$ produces metabolite $m$, $\xi_i^m=-1$ if reaction $i$ consumes metabolite $m$, and $\xi_i^m=0$ otherwise. We furthermore define $\partial m_{\inn}$ (resp. $\partial m_{\outt}$) as the set of reactions producing (resp. consuming) $m$; likewise, for each reaction $i$, $\partial i_{\inn}$ (resp.\ $\partial i_{\outt}$) will denote the set of its substrates (resp.\ products).

% Reactions are generically divided in three classes. {\it Intake reactions} are defined as having $\xi_i^m=\delta_{m,n}$: these processes serve the purpose of supplying an individual metabolite ($n$ here) to the network. {\it Out-take reactions} are defined as having $\xi_i^m=-\delta_{m,o}$, and remove an individual metabolite ($o$ here) from the network. Finally, {\it core reactions} are defined as having a non-zero number of both inputs and outputs.

\begin{figure}
\includegraphics[scale=0.35]{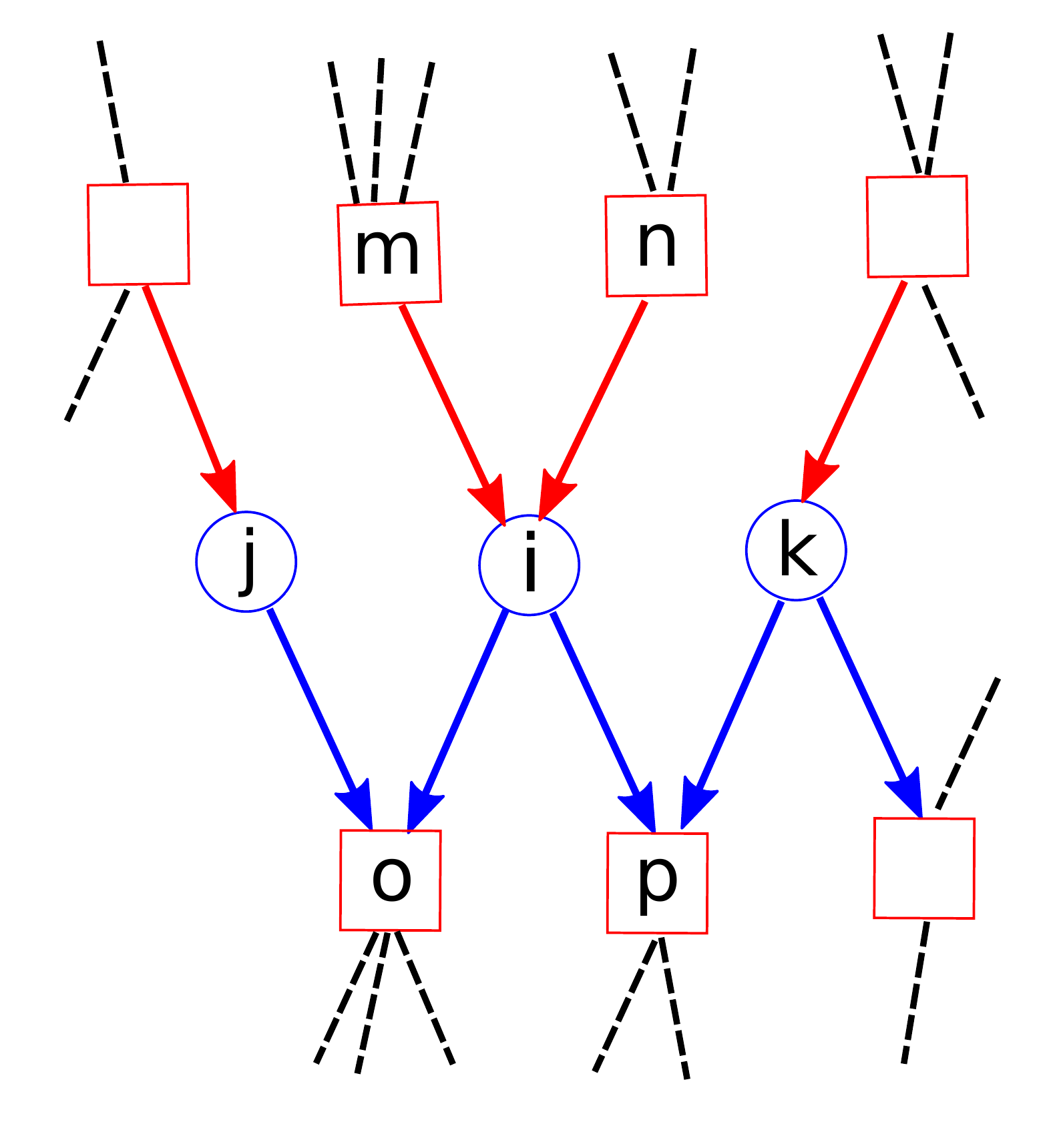}%0.25
\caption{Sketch of a random reaction network of the type discussed in the text. $m,n,o$ and $p$ denote metabolites (squares), $i,j$ and $k$ are instead reactions (circles).  Red (resp. blue) links carry substrate-like (resp. product-like) couplings with $\xi_i^m=-1$ (resp. $\xi_i^p=1$).   \label{bip_sketch}}
\end{figure}

The topology of the RRN is specified by the probability distributions of the degrees of the two node types. For metabolite nodes  we shall assume that the in- and out-degrees $\ell_{\inn}\equiv|\partial m_{\inn}|$ and $\ell_{\outt}\equiv|\partial m_{\outt}|$ are independent random variables, both distributed according to a Poissonian with parameter $\lambda$, i.e.
\begin{equation}
\mathit{D}_M(\ell) = e^{-\lambda}\frac{\lambda^{\ell}}{\ell !}\;.
\label{degree_M}
\end{equation}
Metabolites having $(\ell_{\inn},\ell_{\outt})=(0,0)$ are disconnected from the network and will be ignored in what follows. We shall generically assume that $\lambda\geq \lambda_p=1$ (the percolation threshold), ensuring the existence of a `giant' connected subgraph. Metabolites with $(\ell_{\inn},\ell_{\outt})=(0,\ell\ge1)$ represent the substrates that the reaction network derives from the environment (the `nutrients'), whereas metabolites with $(\ell_{\inn},\ell_{\outt})=(\ell\ge1,0)$ will be considered to be the final products or sinks (e.g. excreted compounds or molecules that are employed in processes other than chemical reactions) of the network. The fraction of such `leaves' (nutrients or sinks) is given by $e^{-\lambda}(1-e^{-\lambda})\simeq e^{-\lambda}$ for large enough $\lambda$. Likewise, for reaction nodes, the quantities $|\partial i_{\inn}|\equiv d_\inn $ and $|\partial i_{\outt}|\equiv d_{\outt}$ will be assumed to be independent random variables, both distributed according to
\begin{equation}
\mathit{D}_R(d)= q\delta_{d,2}+(1-q)\delta_{d,1}\;,
\label{degree_R}
\end{equation} 
where $0\leq q\leq 1$ is a parameter. In other words, reactions can be of four different types according to their in- and out-degrees ($(d_{\inn},d_{\outt}) \in \{(1,1),(1,2),(2,1),(2,2)\}$) and $q$ weights the relative number of bi-component reactions (as inputs, outputs or both). The only structural control parameters that we shall use in the following are the mean degrees of metabolites ($\lambda$) and of reactions ($q$). A sketch of the network is given in Fig. \ref{bip_sketch}.

\subsection{Constraints: Hard versus Soft Mass Balance}

In order to define a CSP embodying realistic operational constraints, we focus on the characterization of the NESS induced by non-zero in- and out-fluxes of nutrients and sinks, respectively \cite{Beard_2008,ksch}, following two different (but related) schemes. In Flux-Balance-Analysis (FBA) it is assumed that fluxes in NESS ensure mass balance at each metabolite node in the network \cite{Kauffman_2003}. (We do not consider here the optimization schemes that are typically coupled to such constraints in biological implementations of FBA \cite{Palsson_book,Feist}.) If we denote by $J_i$ the flux of reaction $i$ (with $J_i\geq 0$ for an irreversible reaction), this amounts to solving the system
\begin{equation}\label{mbe}
\sum_{i=1}^N \sigma_i^mJ_i=0~~,\qquad \forall m\in\{1,\ldots,M\}~~,
\end{equation} 
where $\sigma_i^m$ is the stoichiometric coefficient of metabolite $m$ in reaction $i$ (such that $\text{sgn}(\sigma_i^m)=\xi_i^m$). One easily understands that the above conditions are equivalent to Kirchhoff's node laws for the flow of matter through metabolite nodes and describe NESS with constant (time-independent) levels for each metabolite. A soft version of this model \cite{DeMartino_theo,dmm} assumes instead that intracellular concentrations may be allowed to increase linearly over time at constant rate, e.g. because some metabolites have to be available for processes outside of metabolism strictly defined. This simply leads to replacing (\ref{mbe}) with
\begin{equation}\label{vn}
\sum_{i=1}^N \sigma_i^mJ_i\geq 0~~,\qquad \forall m\in\{1,\ldots,M\}~~.
\end{equation} 
More formally, the above conditions can be seen to derive from Von Neumann's optimal growth scenario \cite{Gale_1989} and provide a useful means of characterizing a reaction network's production capabilities \cite{Imielinski_2006,DeMartino_genes,DeMartino_prodCap}. 

The general problem posed by (\ref{mbe}) and (\ref{vn}) consists, given the matrix $\widehat{\sigma}=\{\sigma_i^m\}$, in retrieving the flux vectors $\mathbf{J}=\{J_i\}$ satisfying the $M$ linear conditions. An interesting feature that is observed in the solutions of the above models is that a sizeable fraction of reactions carries a null flux in each solution \cite{DeMartino_genes,Nishikawa_2008}. This suggests that, to a first approximation, if one is interested in capturing certain aspects of NESS within a coarse-grained description it might suffice to just distinguish, for each reaction, the inactive state from the active one. We shall then introduce, for each reaction, a variable $\nu_i\in\{0,1\}$ (inactive/active). Similarly, we shall link to every metabolite a variable $\mu_m\in\{0,1\}$ that characterizes whether that particular chemical species is available ($\mu_m=1$) or not ($\mu_m=0$) to enzymes that process it. Our next task is to devise Boolean CSPs that embed the basic features underlied by (\ref{mbe}) and (\ref{vn}), respectively.

Starting from (\ref{mbe}), it is simple to understand that a minimal necessary requirement that is encoded in the mass-balance conditions is that, for each metabolite which is produced by an active reaction, there must be at least one active reaction consuming it, and vice-versa. This means that, for each compound $m$, all assignments of $\nu_i$'s are acceptable except those for which all active reactions either produce or consume it. We can therefore define the number of active reactions producing and consuming chemical species $m$ as
\begin{equation}
x_m\equiv \sum_{i\in\partial m_{\inn}} \nu_i\;\quad ~~~~~\text{and}~~~~~
y_m\equiv \sum_{i\in\partial m_{\outt}} \nu_i~~,
\end{equation}
and, in turn, introduce an indicator function $\Gamma_m\equiv \Gamma_m(\mu_m,\{\nu_i\})$ for every $m$ as
\begin{equation}
\label{constraint_FBA}
\Gamma_m = \delta_{\mu_m,0}\delta_{x_m,0} \delta_{y_m,0} + \delta_{\mu_m,1}(1-\delta_{x_m,0}) (1-\delta_{y_m,0})~~.
\end{equation}
Given a configuration $\{\nu_i\}$, metabolite $m$ will be said to be SAT when $\Gamma_m=1$, i.e. when either no reaction in which it is involved is active ($x_m=0$ and $y_m=0$) and the metabolite is unavailable ($\mu_m=0$), or when the metabolite is available ($\mu_m=1$) and at least one reaction produces it ($x_m>0$) and at least one reaction consumes it ($y_m>0$). Similarly, we define a reaction to be SAT when the indicator function $\Delta_i\equiv \Delta_i(\nu_i,\{\mu_m\})$, given by
\begin{equation}
\Delta_i =\delta_{\nu_i,0}+\delta_{\nu_i,1}\prod_{m \in \partial i}\mu_m\;,
\label{constraint_rea_tmp}
\end{equation}
with $\partial i=\partial i_\inn \cup \partial i_\outt$, equals 1. That is, $i$ can be active only if all its neighbouring metabolites  (including both substrates and products) are available. We note that (\ref{constraint_rea_tmp}) can actually be re-cast as
\begin{equation}
\Delta_i =\delta_{\nu_i,0}+\delta_{\nu_i,1}\prod_{m \in \partial i_{\inn}}\mu_m\;,
\label{constraint_rea}
\end{equation}
according to which $i$ can be active only if all of its inputs are available: it is indeed clear that if a reaction is active but one (say) of its products is unavailable, then the constraint imposed on the metabolite will either be violated or force that metabolite to become available. Notice that $\Delta_i=1$ does not imply that $i$ is active when all of its substrates are available. 

The CSP corresponding to (\ref{mbe}) can then be formulated as follows: {\it find a non-trivial assignment of $\nu_i$'s ($\nu_i$'s not all zero) such that all reactions and all metabolites are SAT, i.e. $\Gamma_m=1~\forall m$ and $\Delta_i=1~\forall i$, with $\Gamma_m$ and $\Delta_i$ given by (\ref{constraint_FBA}) and (\ref{constraint_rea}), respectively}. We shall call this CSP Hard Mass Balance, or Hard-MB for brevity.

In order to get a Boolean representation of (\ref{vn}), we note that the main difference between this case and that of (\ref{mbe}) is that, because of the soft constraint, it is no longer necessary that production fluxes are balanced by consumption fluxes. Therefore, while constraint (\ref{constraint_rea}) remains valid, (\ref{constraint_FBA}) has to be replaced by
\begin{equation}\label{gammam}
\Gamma_m = \delta_{\mu_m,0}\delta_{x_m,0} \delta_{y_m,0} + \delta_{\mu_m,1}(1-\delta_{x_m,0})\;.
\end{equation}
In other terms, metabolite $m$ can be available as soon as at least one reaction producing it is active ($x_m>0$). It is convenient to re-write $\Gamma_m$ for this case as
\begin{gather}
%\Delta_i =\delta_{\nu_i,0}+\delta_{\nu_i,1}\prod_{m \in \partial i_{in}}\mu_m\;, \label{constraint_rea_VN} \\
\Gamma_m = \delta_{\mu_m,0}\delta_{x_m,0} + \delta_{\mu_m,1}(1-\delta_{x_m,0})\;, 
\label{constraint_VN}
\end{gather}
so that the constraint at each metabolite node only includes incoming degrees of freedom, making the directionality inherent in the corresponding CSP explicit. It is straightforward to see that  (\ref{gammam}) or (\ref{constraint_VN}), together with (\ref{constraint_rea}), which retains validity, return the same configurations.  

The CSP corresponding to (\ref{vn}) is then the following: {\it find a non-trivial assignment of $\nu_i$'s such that all reactions and all metabolites are SAT, i.e. $\Gamma_m=1~\forall m$ and $\Delta_i=1~\forall i$, with $\Gamma_m$ and $\Delta_i$ given by (\ref{constraint_VN}) and (\ref{constraint_rea}), respectively}. We shall call this CSP Soft Mass Balance, or Soft-MB for brevity.

Note that the constraints behind the two problems can be written compactly as
\begin{gather}
\Gamma_m = \delta_{\mu_m,0}\delta_{x_m,0} (\delta_{y_m,0})^\alpha + \delta_{\mu_m,1}(1-\delta_{x_m,0}) (1-\delta_{y_m,0})^{\alpha}\\
\Delta_i =\delta_{\nu_i,0}+\delta_{\nu_i,1}\prod_{m \in \partial i_{\inn}}\mu_m
\end{gather}
where $\alpha=1$ for Hard-MB and $\alpha=0$ for Soft-MB. We shall be interested in solutions obtained upon fixing the probability that a nutrient is available, which we denote below as $\rho_{\inn}$. For the moment, no specific assumption will be made on sinks.

\section{Results}

In a nutshell, the above setup aims at retrieving Boolean patterns of activity of reactions (or of metabolite availabilities) induced, on network architectures defined by $q$ and $\lambda$, by the fact that a certain set of metabolites (nutrients) is available from the outset. Ideally, one would like to devise a method to sample configurations $(\boldsymbol{\nu}=\{\nu_i\},\boldsymbol{\mu}=\{\mu_m\})$ with a controlled probability given by
\begin{equation}
P\big(\boldsymbol{\nu},\boldsymbol{\mu}\big) \propto \prod_{m=1}^M \Gamma_m \prod_{i=1}^N \Delta_i e^{\theta \nu_i}\;,
\label{meas}
\end{equation}
which forbids states that don't satisfy all constraints. The `chemical potential' $\theta$ appearing above can be tuned externally in order to concentrate the measure around configurations with a different average fraction $N^{-1} \sum_i \avg{\nu_i}$ of active reactions, where angular brackets represent the average with respect to the measure (\ref{meas}).

\subsection{Soft Mass-Balance}

In order to find the configurations of reaction and metabolite variables that solve the above CSPs one may resort to statistical mechanics techniques. In particular, we have employed a cavity theory to devise a belief propagation/population dynamics algorithm to sample the probability distribution (\ref{meas}). Details about the theory and the algorithms are reported in the Appendix. We shall concentrate here on the  scenario that emerges for different  $q$ and $\lambda$ upon varying two parameters, namely the chemical potential $\theta$ and the probability $\rho_{\inn}$ that nutrients are available. In specific, we have computed the average reaction activity and the average metabolite availability following two protocols:  first, by gradually reducing $\theta$ starting from a large, positive value, and, second, by doing the reverse. Averages obtained in these ways will be denoted, respectively, by $\overline{\avg{\cdots}}_+$ and $\overline{\avg{\cdots}}_-$. These averages (that we call magnetizations, using a statistical physics jargon) need not coincide, in which case the two quantities will display hysteresis when plotted against the chemical potential. Generally, the presence of hysteresis is a main characteristic of a discontinuous (first order) phase transition, while for continuous (second order) ones no hysteresis is observed, as also happens in cases where no phase transition takes place.

\begin{figure}
  \includegraphics[width=17cm]{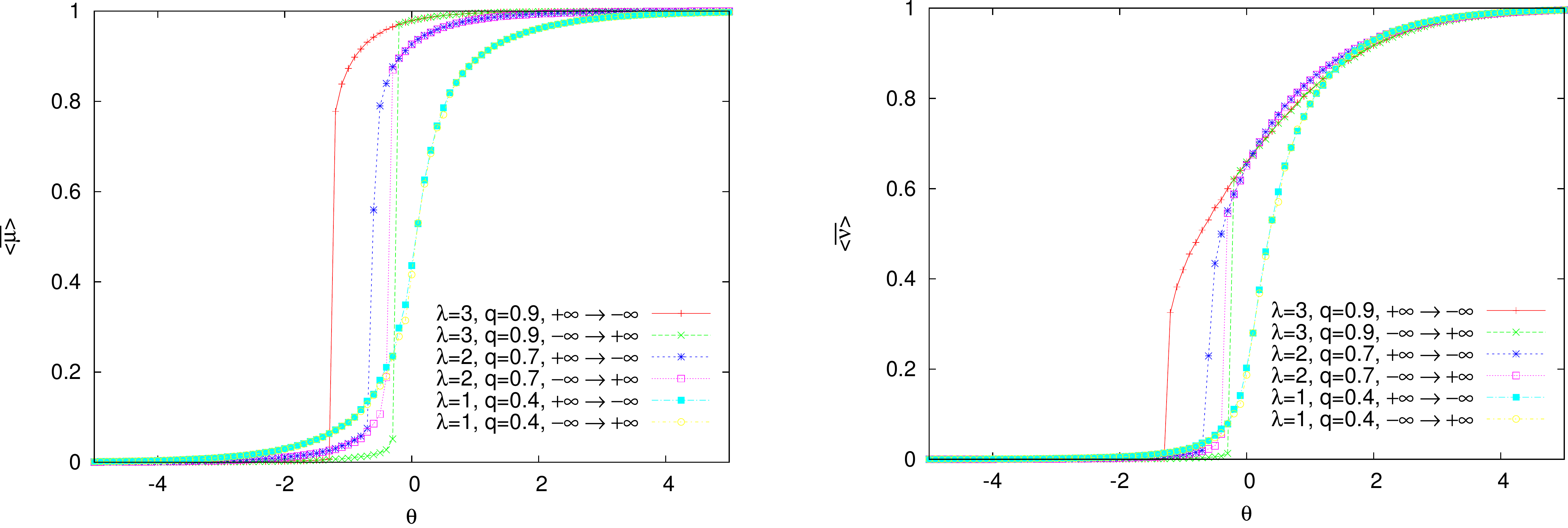}
  \caption{ \label{soft-MB_inTHETA}Soft-MB: behaviour of the average fraction of available metabolites, $\overline{\avg{\mu}}$ (left) and of  the average fraction of active reactions, $\overline{\avg{\nu}}$ (right) versus $\theta$ for different values of the parameters $\lambda$ and $q$ and fixed $\rho_{\inn}=0.5$.} 
\end{figure}

The average fractions of available compounds (metabolites) and active reactions obtained upon varying $\theta$ at fixed $\rho_{\inn}=0.5$ for Soft-MB is displayed in Fig. \ref{soft-MB_inTHETA}.  
One sees that, expectedly, larger values of $\theta$ lead, on average, to larger fractions of available metabolites and of active reactions. For large enough values of $\lambda$ and $q$, however, as the $\overline{\avg{\cdots}}_+$ and $\overline{\avg{\dots}}_-$ averages become steeper functions of $\theta$, the curves obtained by increasing and decreasing $\theta$ no longer coincide. Notice that, while for lower $\lambda$ and $q$ solutions can be found over a broad range of values of the magnetizations, when $\lambda$ and $q$ increase the average metabolite availability seems to concentrate in small ranges close to the extremes 0 and 1, distinguishing solutions with few available metabolites from solutions with a large fraction of available compounds. This type of picture is however not observed for reactions (we shall return to this point later on).

\begin{figure}
\centering
\includegraphics[width=17cm]{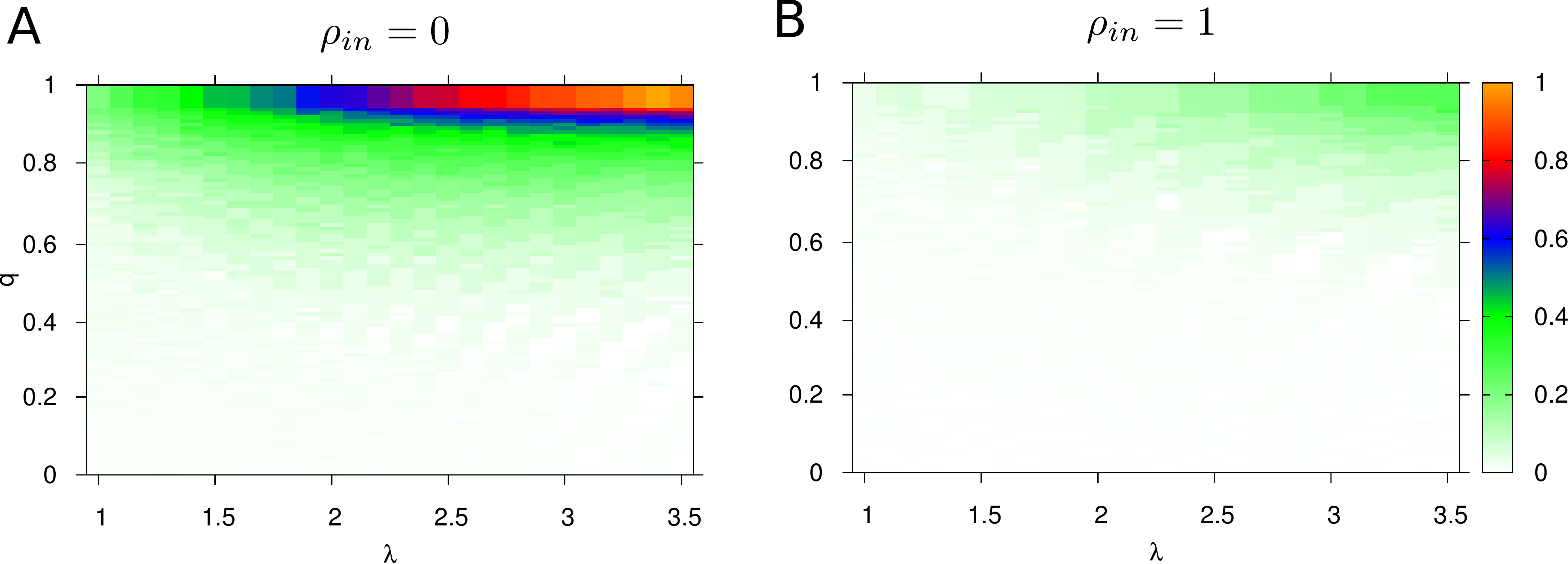}
\caption{Map of the values of $\Delta\mu$ (normalized by the same maximum: 4.852) for the Soft-MB problem in the $(\lambda,q)$ plane. The spacing in $q$ is equal to $0.01$, while it is $0.1$ in $\lambda$. A) $\rho_{\inn}=0$; B) $\rho_{\inn}=1$.}
  \label{phase_diagram_soft-MB}
\end{figure}

A simple way to quantify the onset of hysteresis is by measuring the quantity (we focus for simplicity on metabolites)
\begin{gather}
\Delta \mu =\int_{-\infty}^{+\infty} \left( \overline{\avg{\mu}}_+- \overline{\avg{\mu}}_- \right) d\theta ~~,
\end{gather}
which vanishes when $\overline{\avg{\mu}}_+=\overline{\avg{\mu}}_-$ and generically differs from 0 in presence of hysteresis. A map of the values of $\Delta\mu$ in the parameter space $(\lambda, q)$ is presented in Figure \ref{phase_diagram_soft-MB} for the limiting choices $\rho_\inn=1$ and $\rho_\inn=0$.

While hysteretic behaviour can be found practically all throughout the $(\lambda,q)$ plane, it becomes stronger at high enough $\lambda$ and $q$, where an abrupt jump in the magnetizations takes place. The presence of such a large hysteresis, and the coexistence of low and high magnetization solutions, signal a non trivial structure in the space of solutions to the CSP. Such a non-trivial structure appears also in many other well-known CSP, as the random k-XORSAT \cite{xorsat} and random k-SAT \cite{ksatScience,ksatJSTAT}, and is the origin of the onset of long range correlations, that have important consequences on the behavior of searching algorithms \cite{PNAS}.
Away from the hysteretic portion, $\overline{\avg{\mu}}_+$ and $\overline{\avg{\mu}}_-$ vary smoothly with $\theta$, allowing one to sample easily solutions with any magnetization not in the jump.

On the other hand, the overall structure of the solutions (in terms of $\Delta\mu$) appears to vary weakly with $\rho_\inn$. This strongly suggests that main observed effects (e.g. the jump and the hysteresis) are essentially due to topology of the network, rather than to the boundary conditions.
It is interesting to observe that the hysteretic region shrinks as $\rho_\inn$ increases, suggesting that, within the constraints imposed by Soft-MB, a larger repertoire of available nutrients stabilizes the output by allowing to achieve higher values of the magnetization for smaller values of $\theta$.

\subsection{Hard Mass-Balance}

\begin{figure}
\centering
\includegraphics[width=17cm]{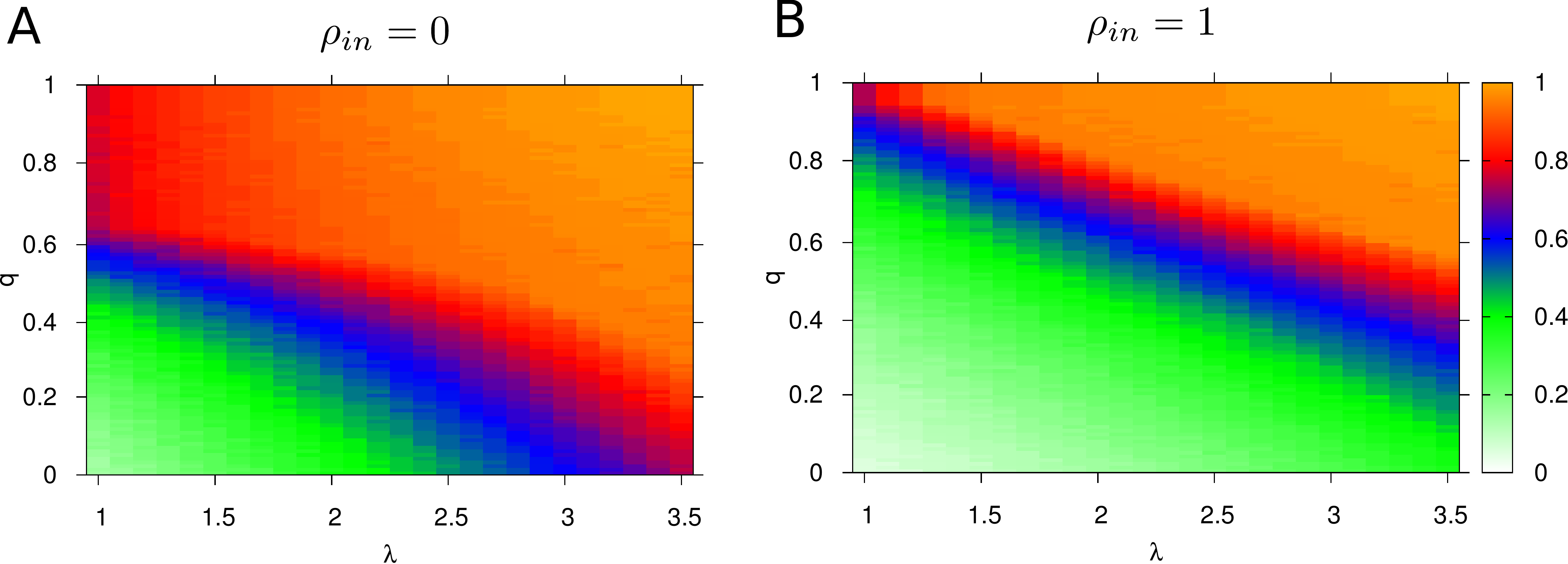}
\caption{\label{phase_diagram_HMB}Map of the values of $\Delta\mu$ (normalized by the same maximum: 12.643) for the Hard-MB problem in the $(\lambda,q)$ plane.  The spacing in $q$ is equal to $0.01$, while it is $0.1$ in $\lambda$. A) $\rho_{\inn}=0$; B) $\rho_{\inn}=1$.}
\end{figure}

The $\Delta \mu$-map for the Hard-MB case is displayed in Figure \ref{phase_diagram_HMB}.
In contrast with the Soft-MB case, Hard-MB solutions display strong hysteresis for all choices of $\lambda$, $q$ and $\rho_\inn$. Furthermore, comparing the results at $\rho_{\inn}=0$ and $\rho_{\inn}=1$, it is clearly seen that, again, changing $\rho_\inn$ (i.e. increasing the number of available nutrients) has little influence on the overall structure of the phase space. Rather, its main effect is that of reducing the magnitude of hysteresis cycles. It is interesting to note that the maximum value of $\Delta \mu$ in Hard-MB is more than double than the one in Soft-MB.

\begin{figure}
\includegraphics[width=17cm]{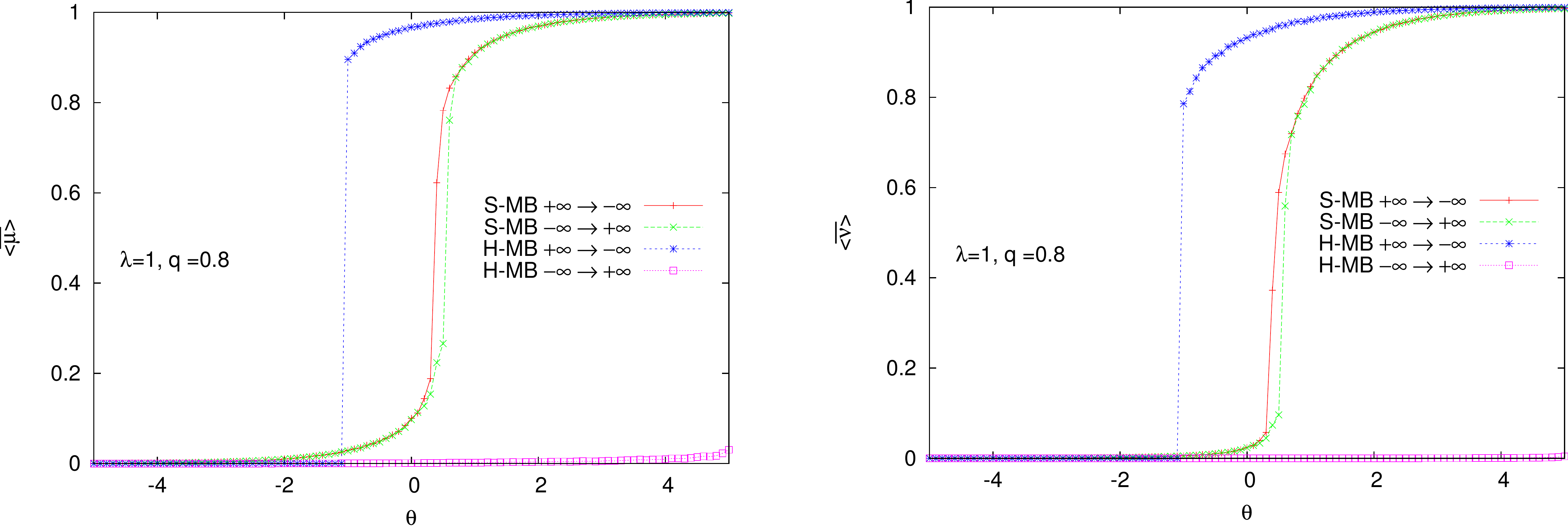}
\caption{Behaviour of $\overline{\avg{\mu}}_+$ and $\overline{\avg{\mu}}_-$ (left) and $\overline{\avg{\nu}}_+$ and $\overline{\avg{\nu}}_-$ (right) as functions of $\theta$ at $\lambda=1$, $q=0.8$ and $\rho_{\inn}=0.5$ for the Soft- and Hard-MB problems.}
\label{VN_FBA_inTHETA_l1}
\end{figure}

\begin{figure}
\includegraphics[width=17cm]{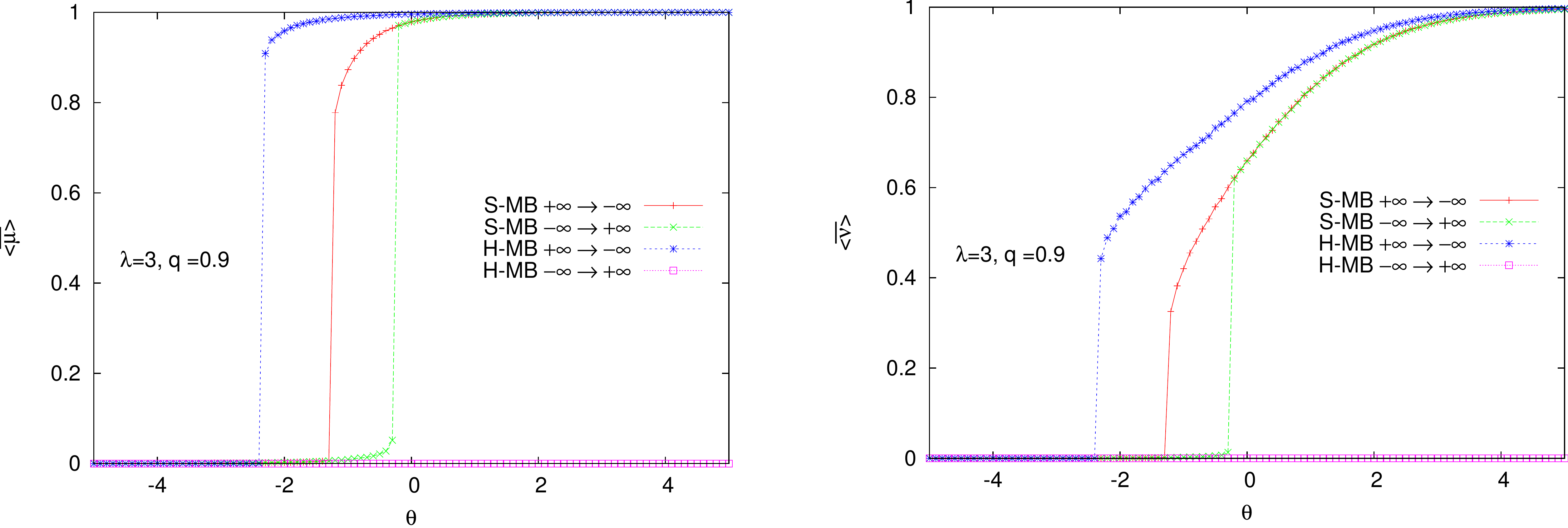}
\caption{Behaviour of $\overline{\avg{\mu}}_+$ and $\overline{\avg{\mu}}_-$ (left) and $\overline{\avg{\nu}}_+$ and $\overline{\avg{\nu}}_-$ (right) as functions of $\theta$  at $\lambda=3$, $q=0.9$ and $\rho_{\inn}=0.5$ for the Soft- and Hard-MB problems.}
\label{VN_FBA_inTHETA_l3}
\end{figure}

The presence of strong hysteresis markedly distinguishes the solution space of the two CSP problems. A comparison between the behaviour of the magnetization obtained in the Soft- and Hard-MB cases for selected parameter values is displayed in Figs \ref{VN_FBA_inTHETA_l1} and \ref{VN_FBA_inTHETA_l3}.
In first place, one sees that the limiting value of the average magnetization for $\theta\to\pm\infty$ in the Hard-MB problem is identical to that of the Soft-MB problem, suggesting that in specific cases the Hard-MB CSP may actually acquire a strong directional nature (like the Soft-MB case), despite the fact that in Hard-MB substrates and products are highly correlated between each other. Secondly, the increasing-$\theta$ protocol appears to be unable to identify active solutions in the Hard-MB case, suggesting that the Hard-MB constraints bias solutions towards activating a large fraction of metabolite nodes.
In the Hard-MB case, it seems that it is possible to start from the all-on configuration and gradually switch off the network, but it is very difficult to switch on part of the network starting from the all-off configuration: for this reason the all-off solution is very stable in the Hard-MB case.

\begin{figure}
\includegraphics[width=17cm]{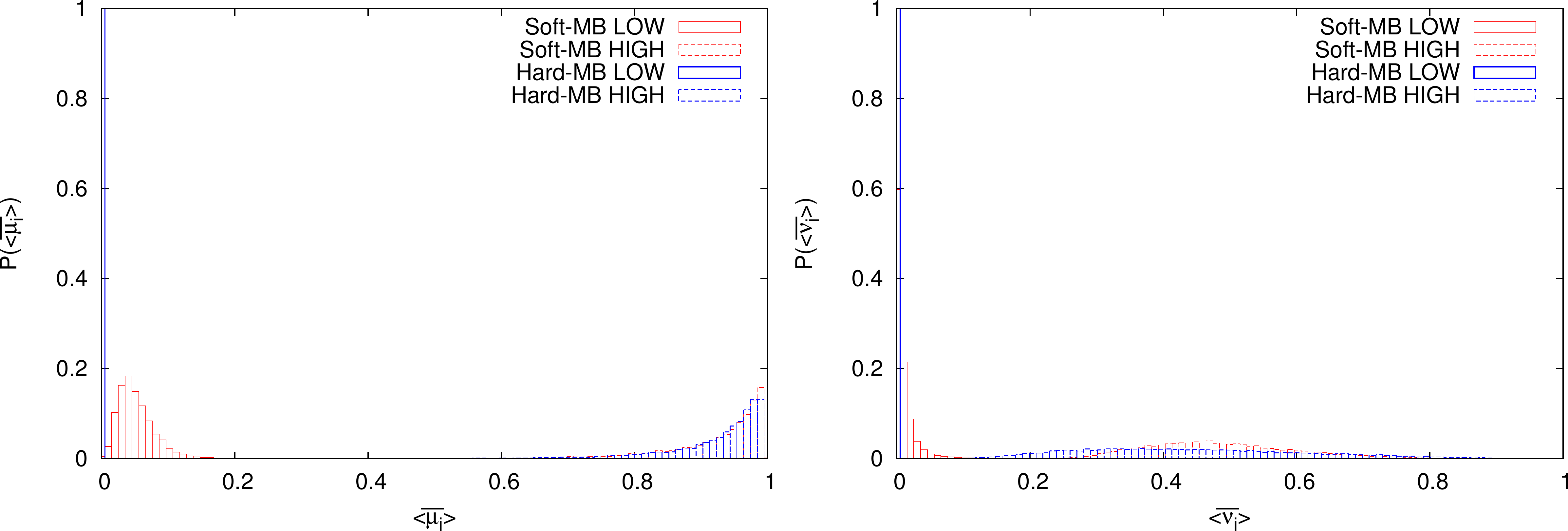}%0.27
\caption{\label{VN_FBA_histogram_mu}Histogram of the values of the average availability of metabolites (left) and reactions (right) for $\lambda=3$, $q=0.8$ and $\rho_{\inn}=0.5$. The value of $\theta$ has been chosen for both CSPs at the transition, so that both high and low values of $\overline{\avg{\mu}}$ and $\overline{\avg{\nu}}$ are possible. In specific, the $\theta$ values for HIGH solutions correspond to $(\overline{\avg{\mu}},\overline{\avg{\nu}}) \simeq (0.92,0.5)$ for both CSPs, while those for LOW solutions corresponds to $(\overline{\avg{\mu}},\overline{\avg{\nu}}) \simeq (0,0)$ for Hard-MB and $(\overline{\avg{\mu}},\overline{\avg{\nu}}) \simeq (0.06,0.014)$ for Soft-MB.}
\end{figure}

A more quantitative view of this is given in Fig. \ref{VN_FBA_histogram_mu}, where we display the  distribution of values of the magnetization for metabolites and reactions obtained for a value of $\theta$ at the transition, where the difference in behaviour between Soft- and Hard-MB is more striking.

From the distribution of metabolite availabilities one clearly sees that, generically, fluctuations are larger in Soft-MB than in Hard-MB, implying that, while Soft-MB sustains non-trivial solutions over a wide range of values of the magnetizations, Hard-MB only admits solutions with a large and tightly constrained value of the average metabolite availability. Interestingly, the overall structure of the distributions changes when one considers reactions, for which both Soft- and Hard-MB can lead a large variability (much larger, in turn, than what occurs for metabolites). This is consistent with our constraints, which do not impose to activate a reaction even when all of its neighbouring metabolites are available. Note that both for reactions and metabolites Soft-MB allows for solutions with very low magnetization that are generically absent in Hard-MB.

\begin{figure}
\centering
\includegraphics[width=17cm]{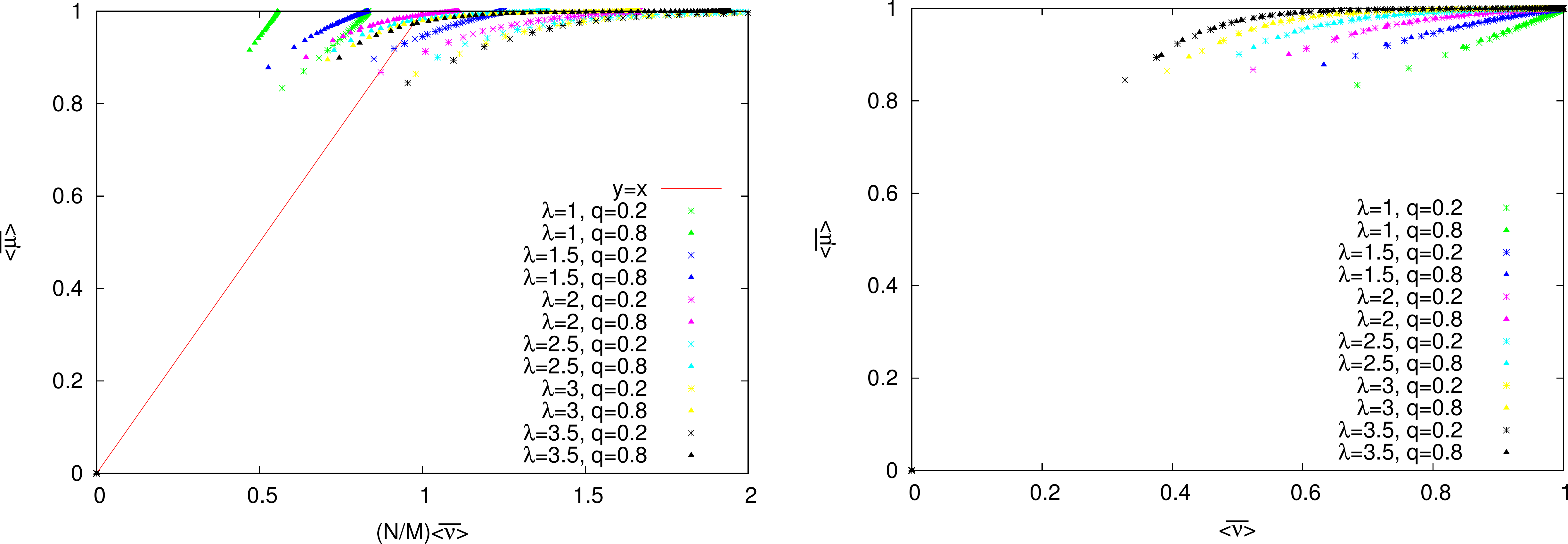}%0.27
\caption{\label{meta_rea_size_comparison}Left: behaviour of $\overline{\avg{\mu}}$ versus $\frac{N}{M}\overline{\avg{\nu}}$ (left) and versus $\overline{\avg{\nu}}$ (right) for $\lambda$ and $q$ as displayed in the legend. In each dataset $\theta$ increases from left to right.}
\end{figure}

Finally, we notice that not all of the solutions to Hard-MB would be able to carry non-vanishing fluxes in the linear problem defined by (\ref{mbe}), which is only possible if the number of available metabolites does not exceed that of active reactions. To see this, one can compare the quantities $M\overline{\avg{\mu}}$ and $N\overline{\avg{\nu}}$, see Figure \ref{meta_rea_size_comparison} (left panel), which are respectively the number of equations and the number of unknowns in the FBA problem. 

It is clear that only for sufficiently large values of $\lambda$ will Boolean configurations correspond to realizable flux states in FBA. This confirms the intuition that redundant network structures (larger $\lambda$'s) confer flexibility (i.e. the possibility of operating in different states) to a reaction network.
What looks counterintuitive in Figure \ref{meta_rea_size_comparison} (left panel) is that small $q$ values are also to be preferred. An explanation to this fact can be obtained by plotting $\overline{\avg{\mu}}$ versus $\overline{\avg{\nu}}$ (right panel in Figure \ref{meta_rea_size_comparison}), and noticing that data with different $q$ values fall on the same curve \footnote{The reasons why curves in the right panel in Figure \ref{meta_rea_size_comparison} depend on $\lambda$ but not  on $q$ rely on a particular property of a random graph, whose detailed derivation is outside the scope of the present paper. We leave this fact as an observation in the numerical data.}
Since the data in the left panel of Figure \ref{meta_rea_size_comparison} are obtained by multiplying the $x$ values in the right panel by $N/M=\lambda/(1+q)$, large $q$ data are more keen to cross the line at the boundary of the feasible solutions region.

In the right panel of Figure \ref{meta_rea_size_comparison} we also notice that $\overline{\avg{\mu}}$ spans a rather limited range (roughly $0.8 < \overline{\avg{\mu}} \le 1$) which is mostly independent on the topology (i.e., on $\lambda$), while the range of valid $\overline{\avg{\nu}}$ values becomes very broad for redundant networks (i.e., for large values of $\lambda$). In other words, solutions to the Boolean constrained problem on a RRN do exist only if a very large fraction of metabolites are present, while the fraction of active reactions can be made small only if the topology is redundant enough.

\section{Outlook}

In this work we have defined and studied a class of random Boolean CSPs representing minimal feasibility and operational constraints for reaction networks. We focused specifically on two sets of conditions: (a) mass balance for metabolites, corresponding to NESS with time-independent concentrations, and (b) mass unbalance allowing for a global net production of chemical species, corresponding to NESS with distinct metabolic output profiles linearly dependent on time. In both cases, we have been interested in computing statistical properties of the solutions induced by exchanges of the network with the environment, modelled by allowing for a fraction of chemical species (with precise topological characteristics) to be available from the outset, either as nutrients (basic inputs) or as sinks (basic outputs) of metabolism. We have displayed results obtained via a cavity-derived algorithm, whose theory is fully described in the Appendix. 

A non-trivial dependence of the `magnetizations' (i.e. of the average reaction activity and the average metabolite availability) on the chemical potential has been uncovered, characterized by hysteresis and, hence, by a first-order transition (in $\theta$). While this phenomenology is present in both models, it is more marked in the Hard-MB case. Remarkably, the overall structure of the solution space (in terms of hysteresis) appears to be only weakly dependent on the value of $\rho_\inn$, characterizing the size of the repertoire of available nutrients, suggesting a considerable robustness with respect to environmental changes, at least in the case of RRNs.

We argue here that many properties observed in this paper are typical of random networks, which have very peculiar topological properties: above the percolation threshold a random network is made of a unique connected component, having no modularity. In other words, the fact that neighbours in a RRN are chosen at random does not allow the network to have correlated structures at short scales (the so-called modules), and in this sense a random network is very different from real ones. On a modular network we expect the first-order-like transition to occur in each module separately, thus producing a more complicates pattern when the chemical potential $\theta$ is varied.
Nevertheless in this work we are interested in the influence of the constraints on the exploration of the fixed point.

The method developed here will be employed in the analysis of the feasible patterns of activity of real metabolic networks. A (minor) adjustment required in order to port the tools described here to a realistic setting concerns the fact that, in physiological conditions, reactions may be reversible. It is possible to account for reversibility in the RRN model described here by simply extending the single reaction state space from $\{0,1\}$ to $\{0,\pm 1\}$. The theory carried out in the Appendix is indeed very easily generalized to the latter case. While such a generalization does not appear to bring out significant qualitative differences for RRNs, it is crucial to deal with genome-scale reconstructions of cellular metabolic networks. On the other hand, real instances of biological networks require algorithms that are capable of extracting individual solutions besides statistical properties. The methods developed here will serve as the necessary basis for accomplishing such a task.

\acknowledgments This work is supported by the Italian Research Minister through the FIRB Project No. RBFR086NN1XYZ and by the DREAM Seed Project of the Italian Institute of Technology (IIT). The IIT Platform Computation is gratefully acknowledged.

\bibliographystyle{unsrt}
\bibliography{biblio}

\appendix

\section*{Appendix: Cavity approach for Soft- and Hard-MB problems}

\subsection{Cavity equations}
\label{cavity_derivation}

In general a CSP, as Soft-MB or Hard-MB, can be solved efficiently on random networks by the belief propagation algorithm \cite{Pearl_BP} or equivalently by the replica symmetric cavity method \cite{Parisi_cavity}. In this method, the marginal of a variable is computed by creating a ``cavity'' inside the system, removing a subpart of the network. Thus it is possible to obtain a ``cavity marginal'' and then reintroduce the variables removed. Finally the complete marginal of the variables follows directly from the cavity marginals. 

In this kind of approach the system is represented by a graph made of ``variable'' and ``function'' nodes. 
Variable nodes are both metabolites and reactions, while a function node exists for each constraint.
In the following we will use letters $a,b,..$ for the metabolite constraints and $e,f,..$ for the reaction constraints. Furthermore we introduce the condensed notations: if $a$ is the constraint of metabolite $m$, then $\partial a^R=\partial a \backslash m$ is the set of reactions involving metabolite $m$; if $e$ is the constraint of reaction $i$, then $\partial e^M=\partial e \backslash i$ is the set of metabolites involved in reaction $i$. Moreover by dividing in two groups reactions producing and consuming a given metabolite, we call $\partial a^R_i$ the set of reactions in the same group as $i$, excluding $i$, and $\partial a^R_{\neg i}$ the opposite group.

Using this notation we can rewrite the constraints (\ref{constraint_rea}) and (\ref{constraint_VN}) for the Soft-MB case in a simpler form that will be useful in the computation of the equations of the system:
\begin{gather}
 \Gamma_a(\mu_m,\nu_{\neigRea}) =\delta_{\mu_m,0}\prod_{\neigIn}(1-\nu_j)+\delta_{\mu_m,1}(1-\prod_{\neigIn}(1-\nu_j))\;, 
\nonumber\\ \label{constraint_VN_final}\\ \nonumber
 \Delta_e(\mu_{\neigMeta},\nu_i) =\delta_{\nu_i,0}+\delta_{\nu_i,1}\prod_{n \in \partial e^M}\mu_n\;,
\end{gather}
where we have substituted $\delta_{x_m,0}=\prod\limits_{j \in \partial a^R_{in}}(1-\nu_j)$, with $a^R_{in}$ being the set of reactions producing the metabolite whose constraint is $a$.

In the same way, for the Hard-MB case, the constraints in the new notation can be written as:
\begin{gather}
  \Gamma_a(\mu_m,\nu_{\neigRea}) =\delta_{\mu_m,0}\prod_{\neigIn}(1-\nu_j)\prod_{\neigOut}(1-\nu_j)+\delta_{\mu_m,1}(1-\prod_{\neigIn}(1-\nu_j))(1-\prod_{\neigOut}(1-\nu_j))\;, \nonumber\\ \label{constraint_FBA_final} \\ \nonumber
  \Delta_e(\mu_{\neigMeta},\nu_i) =\delta_{\nu_i,0}+\delta_{\nu_i,1}\prod_{n \in \partial e^M}\mu_n\;.
\end{gather}

\begin{figure}
  \centering
  \includegraphics[width=8.5cm]{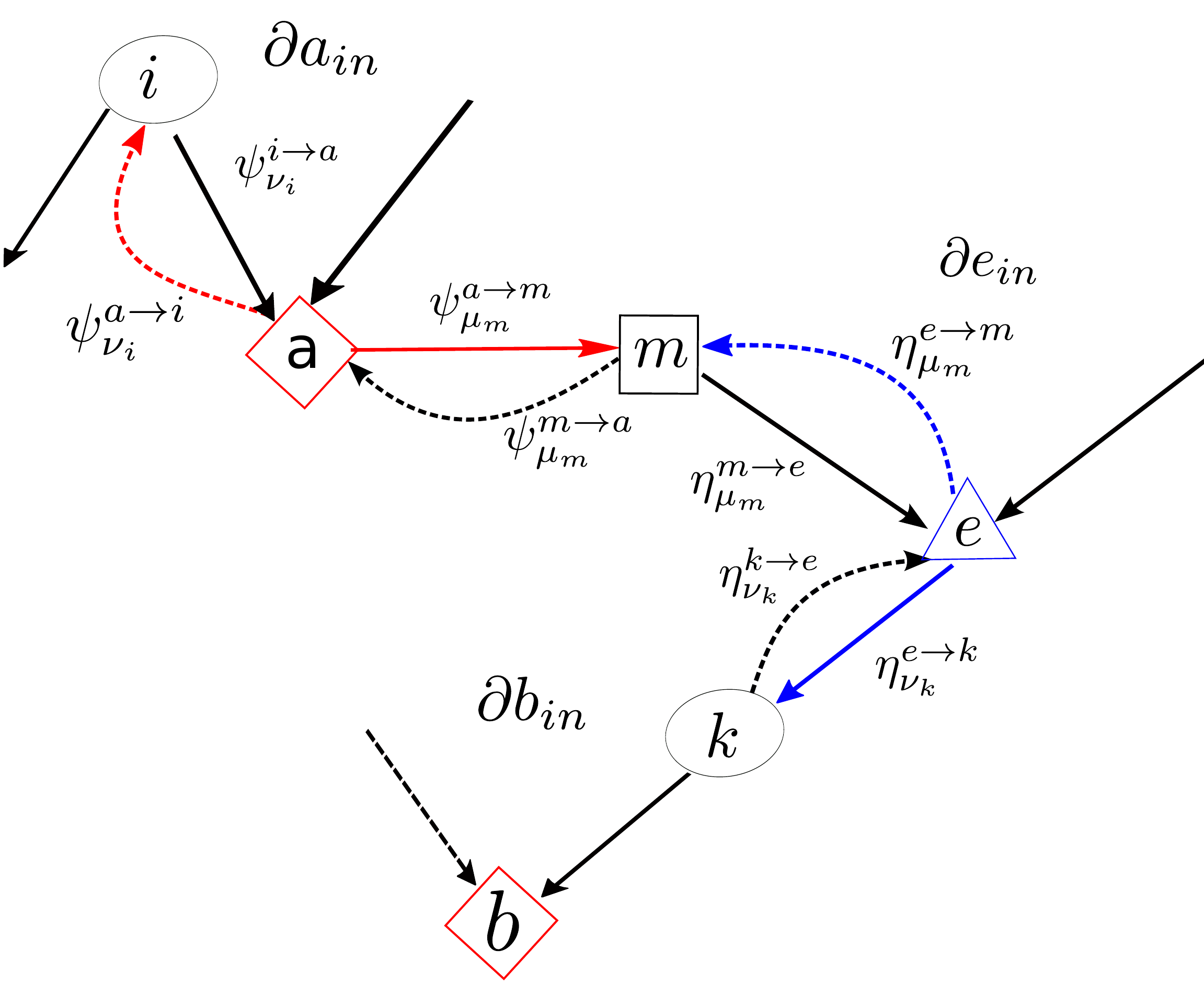}
  \includegraphics[width=8.5cm]{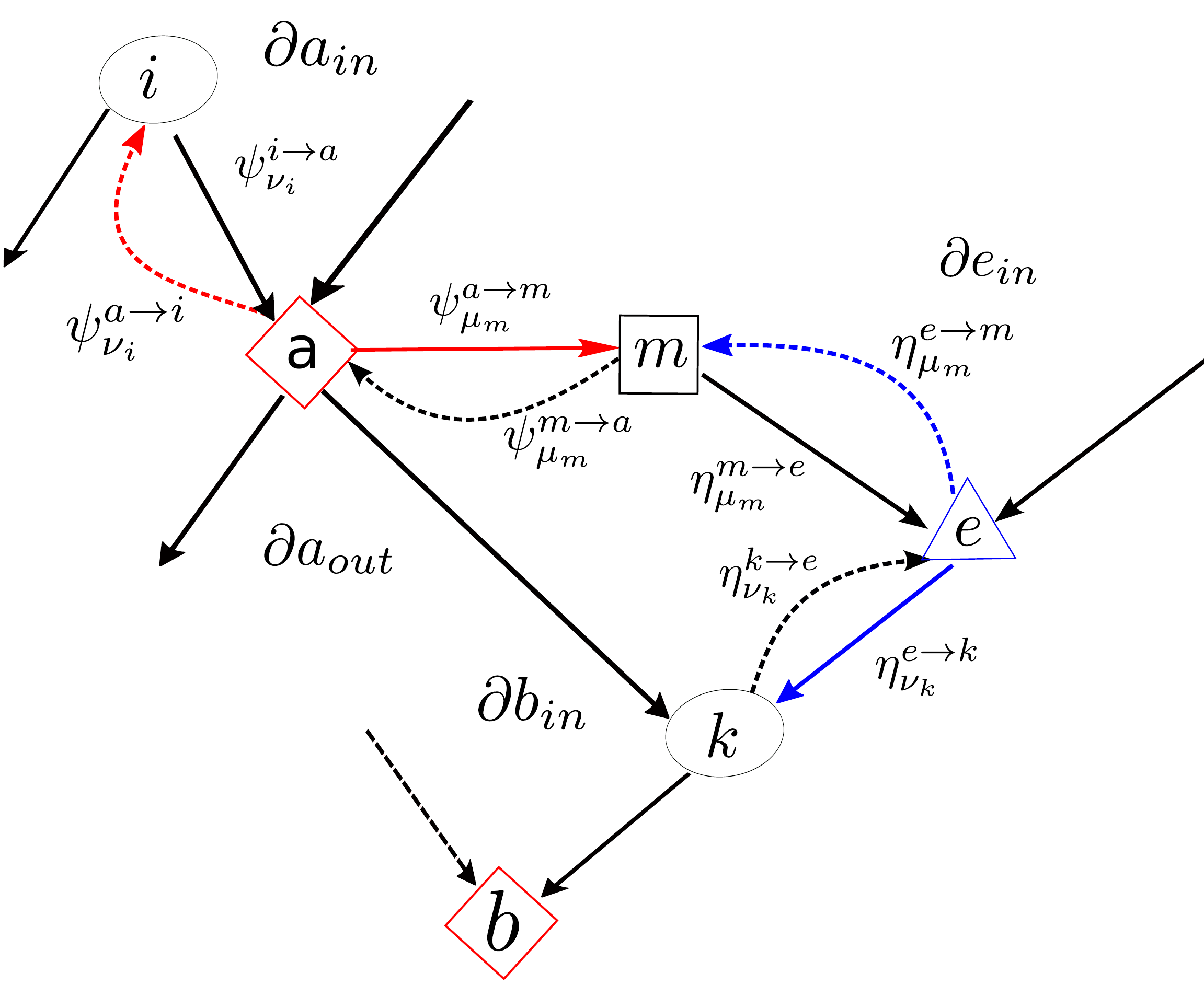}
  \caption{Summary of the cavity method messages for Soft-MB (left) and Hard-MB (right) constraints.   \label{schema_cavity}}
\end{figure}

The representation of the graph that we obtain with the variables and the functions nodes is given in Figure \ref{schema_cavity} for Soft-MB and Hard-MB. 
As we can see the main difference between the two approaches is that the constraint $a$ of the metabolite $m$ changes its form. 

From an algorithmic point of view, it is possible to search for the solutions to the cavity equations by an iterative procedure, in which ``messages'' are exchanged between variable and function nodes. For the two CSP defined in the present work, eight type of messages are required:  $\mess{\psi}{a}{m}_{\mu_m}$, $\mess{\psi}{m}{a}_{\mu_m}$, $\mess{\eta}{e}{m}_{\mu_m}$, $\mess{\eta}{m}{e}_{\mu_m}$, $\mess{\psi}{a}{i}_{\nu_i}$, $\mess{\psi}{i}{a}_{\nu_i}$, $\mess{\eta}{e}{i}_{\nu_i}$, $\mess{\eta}{i}{e}_{\nu_i}$. Each message represents the belief that a variable (function) has about its neighbouring function (variable) state. The messages can be divided in two classes: from function nodes to variable nodes and from variable nodes to function nodes. The first class of messages, e.g. $\mess{\psi}{a}{m}_{\mu_m}$, is the probability that metabolite $m$ is in state $\mu_m$ when there is only the function metabolite $a$. While the second class, e.g. $\mess{\psi}{m}{a}_{\mu_m}$, is the probability that metabolite $m$ is in state $\mu_m$ when the edge $(am)$ is not present. 

Introducing a parameter $\alpha$ to interpolate between Soft-MB ($\alpha=0$) and Hard-MB constraints ($\alpha=1$), the equations to be satisfied by the messages in the two CSPs under study can be written as follows:
\begin{align*}
 & \begin{cases}
    \mess{\psi}{m}{a}_{\mu_m}=\prod\limits_{f \in \neigRea[m]}\mess{\eta}{f}{m}_{\mu_m}/\Z{m}{a} \\ \\
    \mess{\psi}{a}{m}_{\mu_m}=\sum\limits_{\{\nu_j\}}\Gamma(\mu_m,\nu_{\neigRea})\prod\limits_{k\in \neigRea}\mess{\psi}{k}{a}_{\nu_k}/\Z{a}{m}
  \end{cases}
  \\ \nonumber \\
&  \begin{cases}
    \mess{\psi}{i}{a}_{\nu_i}=\mess{\eta}{e}{i}_{\nu_i}\left( \prod\limits_{b \in \neigMeta[i]_{in} \backslash a}\mess{\psi}{b}{i}_{\nu_i}\right)^\alpha \prod\limits_{b \in \neigMeta[i]_{out} \backslash a}\mess{\psi}{b}{i}_{\nu_i}/\Z{i}{a} \\ \\
    \mess{\psi}{a}{i}_{\nu_i}=\sum\limits_{\{\nu_j\},j\neq i}\sum\limits_{\mu_m}\Gamma(\mu_m,\nu_{\neigRea})\mess{\psi}{m}{a}_{\mu_m}\prod\limits_{k\in \neigRea \backslash i}\mess{\psi}{k}{a}_{\nu_k}/\Z{a}{i}
  \end{cases}
  \\ \nonumber \\
&  \begin{cases}
    \mess{\eta}{i}{e}_{\nu_i}=\left(\prod\limits_{b \in \neigMeta[i]_{in}}\mess{\psi}{b}{i}_{\nu_i}\right)^\alpha\prod\limits_{b \in \neigMeta[i]_{out}}\mess{\psi}{b}{i}_{\nu_i}/\Z{i}{e}  \\ \\
    \mess{\eta}{e}{i}_{\nu_i}=\sum\limits_{\{\mu_n\}}e^{\theta\nu_i}\Delta(\nu_i,\mu_{\neigMeta})\prod\limits_{n \in \neigMeta}\mess{\eta}{n}{e}_{\mu_n}/\Z{e}{i}
  \end{cases} 
  \\ \nonumber \\
&  \begin{cases}
    \mess{\eta}{m}{e}_{\mu_m}=\mess{\psi}{a}{m}_{\mu_m}\prod\limits_{f \in \neigRea[m] \backslash e}\mess{\eta}{f}{m}_{\mu_m}/\Z{m}{e} \\ \\
    \mess{\eta}{e}{m}_{\mu_m}=\sum\limits_{\{\mu_n\},n\neq m}\sum\limits_{\nu_i}e^{\theta\nu_i}\Delta(\mu_{\neigMeta},\nu_i)\mess{\eta}{i}{e}_{\nu_i}\prod\limits_{n\in \neigMeta \backslash m}\mess{\eta}{n}{e}_{\mu_n}/\Z{e}{m}
  \end{cases} 
\end{align*}
The equations we have just written hold in Soft-MB or Hard-MB case, with the difference that in Soft-MB case reaction nodes are connected only to output metabolite functions, while in Hard-MB all metabolite functions are connected to reaction nodes. Another caution we have to take is that the reaction function node is connected only to the input metabolites [see equations (\ref{constraint_VN_final}) and (\ref{constraint_FBA_final})] regardless of the constraints used. Writing explicitly the constraints we can compute the cavity equations obtaining, for the metabolite constraints,
\begin{align*}
&  \begin{cases}
    \mess{\psi}{m}{a}_{\mu_m}=\prod\limits_{f \in \neigRea[m]}\mess{\eta}{f}{m}_{\mu_m}/\Z{m}{a} \\ \\
    \mess{\psi}{a}{m}_{\mu_m}=\left[\delta_{\mu_m,0}\prod\limits_{\neigIn}\mess{\psi}{j}{a}_0\left(\prod\limits_{\neigOut}\mess{\psi}{j}{a}_0\right)^{\alpha}+\delta_{\mu_m,1}\left(1-\prod\limits_{\neigIn}\mess{\psi}{j}{a}_0\right)\left(1-\prod\limits_{\neigOut}\mess{\psi}{j}{a}_0\right)^{\alpha}\right]/\Z{a}{m}
  \end{cases}
  \\ \nonumber \\
  &  \Z{a}{m}=\left(1-\prod_{\neigIn}\mess{\psi}{j}{a}_0\right)\left(1-\prod_{\neigOut}\mess{\psi}{j}{a}_0\right)^{\alpha}+\prod\limits_{\neigIn}\mess{\psi}{j}{a}_0\left(\prod\limits_{\neigOut}\mess{\psi}{j}{a}_0\right)^{\alpha}
  \\ \nonumber \\
&  \begin{cases}
    \mess{\psi}{i}{a}_{\nu_i}=\mess{\eta}{e}{i}_{\nu_i}\left(\prod\limits_{b \in \neigMeta[i]_{in} \backslash a}\mess{\psi}{b}{i}_{\nu_i}\right)^{\alpha}\prod\limits_{b \in \neigMeta[i]_{out} \backslash a}\mess{\psi}{b}{i}_{\nu_i}/\Z{i}{a} \\ \\
    \mess{\psi}{a}{i}_{\nu_i} = \left[\mess{\psi}{m}{a}_0(1-\nu_i)\prod\limits_{\neigIn \backslash i}\mess{\psi}{j}{a}_0\left(\prod\limits_{\neigOut \backslash i}\mess{\psi}{j}{a}_0\right)^{\alpha}+\right.\\ \\
    \qquad\qquad\left. +\mess{\psi}{m}{a}_1\left(1-\prod\limits_{\neigOppositeI}\mess{\psi}{j}{a}_0\right)^{\alpha}\left((1-\prod\limits_{\neigGroupI}\mess{\psi}{j}{a}_0)+\nu_i\prod\limits_{\neigGroupI}\mess{\psi}{j}{a}_0\right)\right]/\Z{a}{i}
  \end{cases}
  \\ \nonumber \\
&\Z{a}{i}=\mess{\psi}{m}{a}_0\prod\limits_{\neigGroupI}\mess{\psi}{j}{a}_0\left(\prod\limits_{\neigOppositeI}\mess{\psi}{j}{a}_0\right)^{\alpha}+ \mess{\psi}{m}{a}_1 \left(1-\prod\limits_{\neigOppositeI}\mess{\psi}{j}{a}_0\right)^{\alpha}\left(2-\prod\limits_{\neigGroupI}\mess{\psi}{j}{a}_0\right)
\end{align*}
and, for the reaction constraints,
\begin{align*}
&  \begin{cases}
    \mess{\eta}{i}{e}_{\nu_i}=\left(\prod\limits_{b \in \neigMeta[i]_{in}}\mess{\psi}{b}{i}_{\nu_i}\right)^{\alpha}\prod\limits_{b \in \neigMeta[i]_{out}}\mess{\psi}{b}{i}_{\nu_i}/\Z{i}{e}  \\ \\
    \mess{\eta}{e}{i}_{\nu_i}=\left[\delta_{\nu_i,0}+e^{\theta}\delta_{\nu_i,1}\prod\limits_{n\in \neigMeta}\mess{\eta}{n}{e}_1\right]/\Z{e}{i} 
  \end{cases} 
  \\ \nonumber \\
& \Z{e}{i}=1+e^{\theta}\prod_{m \in \neigMeta}\mess{\eta}{m}{e}_1
  \\ \nonumber \\
&  \begin{cases}
    \mess{\eta}{m}{e}_{\mu_m}=\mess{\psi}{a}{m}_{\mu_m}\prod\limits_{f \in \neigRea[m] \backslash e}\mess{\eta}{f}{m}_{\mu_m}/\Z{m}{e} \\ \\
    \mess{\eta}{e}{m}_{\mu_m}=\left[\mess{\eta}{i}{e}_0+e^{\theta}\mess{\eta}{i}{e}_1\mu_m\prod\limits_{n\in \neigMeta \backslash m}\mess{\eta}{n}{e}_1\right]/\Z{e}{m}
  \end{cases}
  \\ \nonumber \\
& \Z{e}{m}=2\mess{\eta}{i}{e}_0+e^{\theta}\mess{\eta}{i}{e}_1\prod\limits_{n\in \neigMeta \backslash m}\mess{\eta}{n}{e}_1
\end{align*}

Using these equations, we can iterate until convergence the algorithms presented in Appendix \ref{BP_explanation} and \ref{PopDyn_explanation}, finding solutions that satisfies the constraints and obtaining the cavity marginals. We can then compute the real marginals of the variable nodes as:
\begin{align}
& p(\mu_m)=\mess{\psi}{a}{m}_{\mu_m}\prod\limits_{f \in \neigRea[m]}\mess{\eta}{f}{m}_{\mu_m}/Z^m,
\nonumber \\ \label{proba_rel} \\ \nonumber
& p(\nu_i)=\mess{\eta}{e}{i}_{\nu_i}\left(\prod\limits_{b \in \neigMeta[i]_{in}}\mess{\psi}{b}{i}_{\nu_i}\right)^\alpha\prod\limits_{b \in \neigMeta[i]_{out}}\mess{\psi}{b}{i}_{\nu_i}/Z^i,
\end{align}
where:
\begin{align}
& Z^m=\sum_{\mu_m}\mess{\psi}{a}{m}_{\mu_m}\prod\limits_{f \in \neigRea[m]}\mess{\eta}{f}{m}_{\mu_m}.
\nonumber \\ \label{normalizations} \\ \nonumber
& Z^i=\sum_{\nu_i}\mess{\eta}{e}{i}_{\nu_i}\left(\prod\limits_{b \in \neigMeta[i]_{in}}\mess{\psi}{b}{i}_{\nu_i}\right)^\alpha\prod\limits_{b \in \neigMeta[i]_{out}}\mess{\psi}{b}{i}_{\nu_i},
\end{align}

The main assumption behind the cavity method is that the messages coming from two neighbouring nodes are independent. This clearly depends on the length of loops in the network: if the length of typical loops grows with the system size (as in RRN), then the above assumption can be valid, at least in the thermodynamic limit. As we can see clearly from Figure \ref{schema_cavity} short loops are not present in the Soft-MB, but they arise in the Hard-MB problem. If the assumption breaks down, then message passing algorithms may fail to converge, although it has been observed that also in networks with many short loops (as e.g. regular lattices) it is possible to find solutions to the cavity equations by iterative message passing algorithms \cite{GBP_2D}. In the two CSPs under study we have observed a better convergence of the algorithm for Soft-MB constraints ($\alpha=0$) than for Hard-MB constraints ($\alpha=1$), and to ensure convergence in any case we have reached the Hard-MB limit by varying smoothly the $\alpha$ parameter.

\subsection{Nutrients and Outputs}
\label{inputs_and_outputs}

\begin{figure}[h]
  \centering
  \includegraphics[scale=0.35]{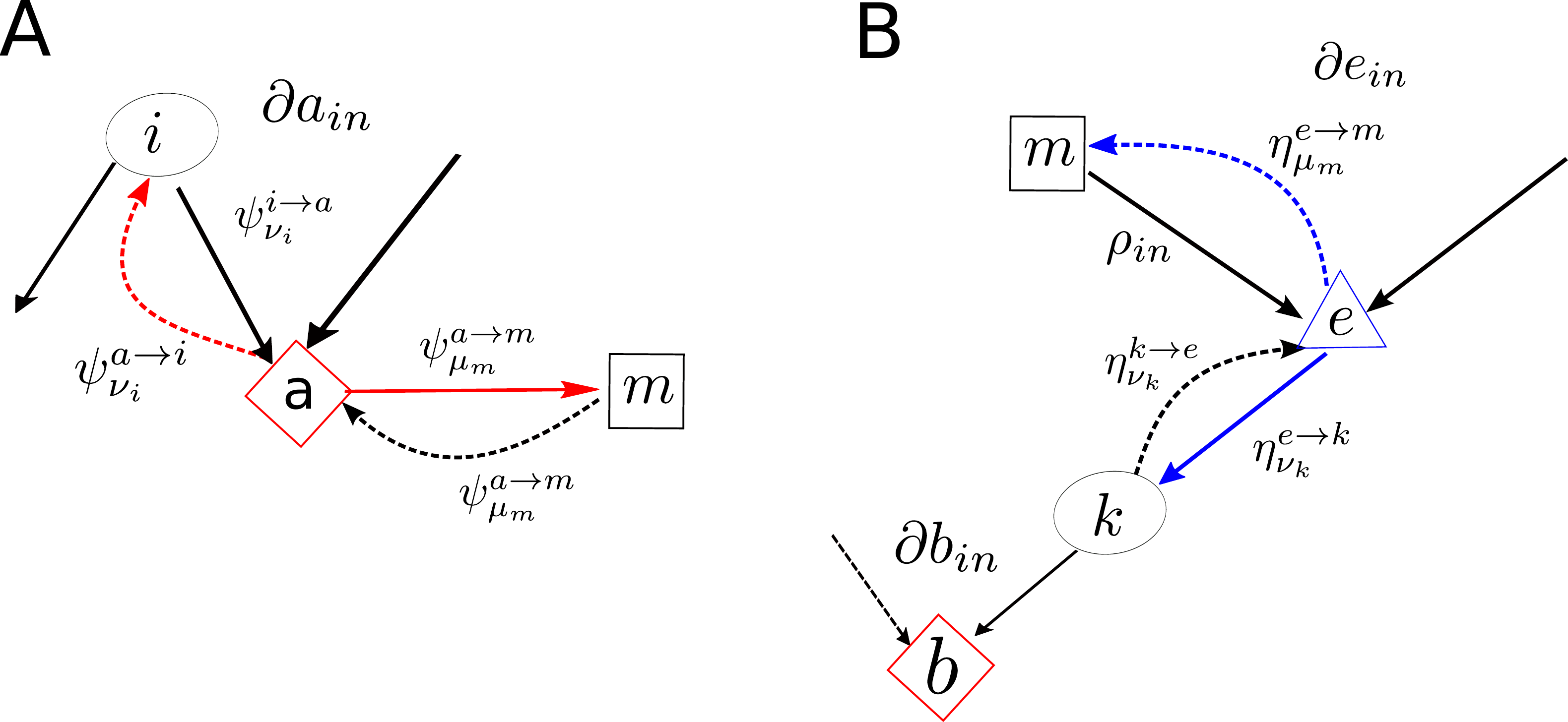}
  \caption{Representation of the external metabolites in our network. \textbf{A} is the product while \textbf{B} is the nutrient.}
  \label{nutrOut_schema}
\end{figure}

How to deal with the nutrients (metabolites with in-degree 0 and out-degree larger than 0) and the outputs (metabolites with in-degree larger than 0 and out degree 0) is probably the trickiest part of the network analysis.  Indeed looking at the cavity equations derived in \ref{cavity_derivation}, we immediately see that nutrients and outputs are automatically switched off because in these cases $\mess{\psi}{a}{m}_{\mu_m}=\delta_{\mu_m,0}$, while in real systems these variables are usually active, as they represent the interaction with the environment. To overcome this limitation we will consider in the following that nutrients are \textit{external} variables fixed by the environment and thus have a probability of being present $\rho_{in}$. Furthermore these variables send a message to the neighbouring reaction-constraint of the type:
\begin{gather}
\mess{\eta}{m}{e}_{\mu_m}=  (1-\rho_{in})\delta_{\mu_m,0} + \rho_{in}\delta_{\mu_m,1}.
\end{gather}
On the other hand the products are \textit{internal} variables with no reaction constraint node associated, a probability of being present $p(\mu_{m})=\mess{\psi}{a}{m}_{\mu_m}$ (taken from \ref{proba_rel}) and send a message:
\begin{gather}
\label{output_message}
\mess{\psi}{m}{a}_{\mu_m}=\mess{\psi}{a}{m}_{\mu_m},
\end{gather}
to the neighbouring metabolite-constraint node. Furthermore the metabolite-constraint has to be a Soft-MB constraint otherwise the outputs would be always off in the Hard-MB case. A schematic view of the form of the network for the external metabolites is presented in Figure \ref{nutrOut_schema}.

\begin{figure}
  \centering
  \includegraphics[scale=0.45]{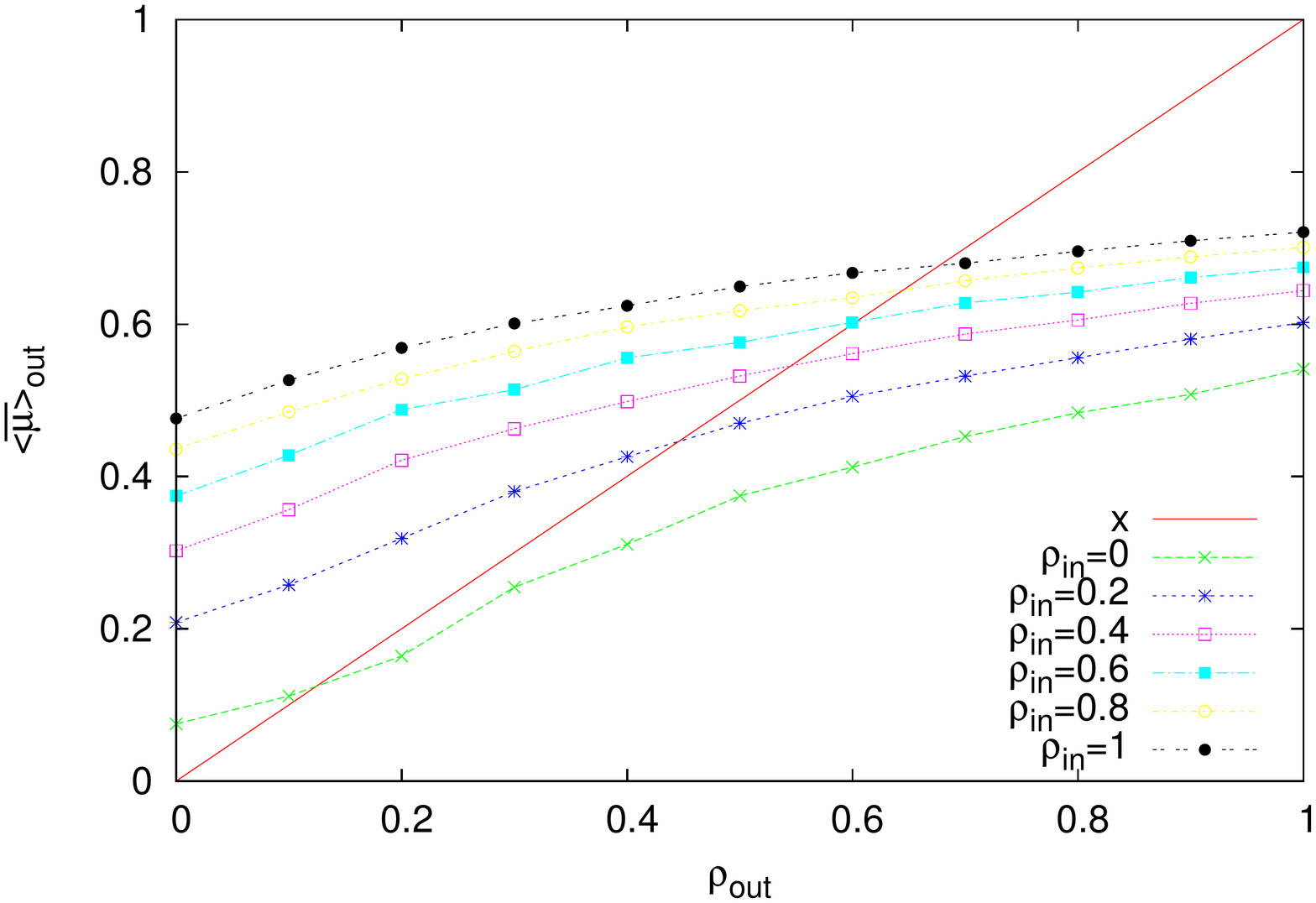}
  \caption{Plot of $\overline{\avg{\mu}}_{out}$ versus $\rho_{out}$ for $q=0.4$, $\lambda=1.5$ and various values of $\rho_{in}$.}
  \label{inRHOOUT}
\end{figure}

In principle for the outputs it is possible to define another parameter $\rho_{out}$ as:
\begin{gather}
\mess{\psi}{m}{a}_{\mu_m}=   (1-\rho_{out})\delta_{\mu_m,0} + \rho_{out}\delta_{\mu_m,1},
\end{gather}
to force the network to switch on a fraction of outputs. Nevertheless it is then required to check at convergence of the algorithm that this value is consistent with the value of $p(\mu_m)$. A simple way to check this is by measuring $\overline{\avg{\mu}_{out}}=\overline{\sum\limits_{i \in outputs} \mu_i/N_{out}}$ and checking if this value is consistent with the value of $\rho_{out}$ given as a parameter. In Figure \ref{inRHOOUT} this check is done for a particular case. In this Figure we see that there is only one value of $\rho_{out}$ consistent with $\overline{\avg{\mu}}_{out}$ (this result holds similarly for Soft-MB and Hard-MB, and for any $q$ and $\lambda$). We then verified that this solution is exactly the same as the one obtained by using equation (\ref{output_message}), hence showing that the parameter $\rho_{out}$ is not necessary to explore all the possible solutions of the outputs.

As a consequence of this setting on the inputs and outputs, the presence of the nutrients is determined by the parameter $\rho_{in}$ while the presence of the outputs is determined at convergence depending on the state of the network.

\subsection{Belief Propagation Algorithm}
\label{BP_explanation}
In order to solve the equations we can run an iterative algorithm until convergence of the value of the messages. Thus the outcome of this algorithm is a set of probabilities of the states in which the variables are. This kind of algorithm is called Belief Propagation (BP). 

The BP algorithm that we used to sample the solutions of the system is the same for Soft-MB and for Hard-MB. In this algorithm, first we generate a graph with a given $q$ and $\lambda$. Then we initialize the messages with a random value and we iterate the equations until convergence. The simplest way to sample the solutions is by fixing one of the two free variables remained: $\theta$ or $\rho_{in}$. By changing $\rho_{in}$ we can see how the configuration of the solutions changes when the nutrients have a probability $\rho_{in}$ of functioning. Whereas by changing $\theta$ we can observe what happens if we constrain the system to switch on (or off) the reactions. Each behaviour is interesting to understand how the system is organized. In each case we will compute the mean over the metabolites,$\avg{\mu}$, and the reactions, $\avg{\nu}$, where $\avg{x}$ is the average over the measure $P(\mu,\nu)$, (\ref{meas}).

\subsection{Population dynamics}
\label{PopDyn_explanation}

BP is an algorithm for inferring the marginal probabilities on a specific graph. However, when one is interested in the behavior of typical samples of the RRNs with given parameters $q$ and $\lambda$, then the equations presented in Appendix \ref{cavity_derivation} can be solved using {\it population dynamics} \cite{Parisi_cavity}. The idea behind this approach is that, instead of computing the messages on a given graph, one considers the probabilities, $P(\psi)$ and $Q(\eta)$, of having a message $\psi$ or $\eta$ in the system. 
Self-consistency equations for these probabilities can be written as follows:
\begin{align}
P(\psi)=E_{\lambda,q}\left[\prod \int d\eta \;Q(\eta)\; d\psi'\; P(\psi') \delta(\psi-F(\eta,\psi'))\right], \\ \nonumber \\
Q(\eta)=E_{\lambda,q}\left[\prod \int d\eta' \;Q(\eta')\; d\psi\; P(\psi) \delta(\eta-G(\eta',\psi))\right],
\end{align}
where the product is over the neighbours and the functions $F(\eta,\psi)$ and $G(\eta,\psi)$ are given by the equations in Appendix \ref{cavity_derivation}.
These population dynamics equations can be solved iteratively and once the fixed point has been reached, averages over the RRN {\it ensemble} can be directly computed.

In the population dynamics algorithm, we start by initializing the system with a random {\it population} of messages and by fixing the parameters of the RRN, $q$ and $\lambda$. Then we iterate using the equations of Appendix \ref{cavity_derivation} where the neighbours are extracted at random, using the distributions (\ref{degree_M}) and (\ref{degree_R}). This is done until convergence of the mean of the messages in the system. At convergence we can compute the mean value of the metabolites, $\overline{\avg{\mu}}$, and reactions, $\overline{\avg{\nu}}$, with respect to the ensemble of RRNs and over the measure (\ref{meas}).

\end{document}